%% file: combi_HE.tex
\documentclass[12pt]{article}

\pdfoutput=1

\usepackage{jheppub}
\usepackage{graphicx,epsfig,amsmath,amssymb}
\usepackage{subcaption}
\usepackage{multirow}
\usepackage{slashed}
\usepackage{xspace}

\newcommand{\als}{\ensuremath{\alpha_s}}

\newcommand{\msbar}{\ensuremath{\overline{\mathrm{MS}}}\xspace}
\newcommand{\mtmsb}{\ensuremath{\overline{m}_t}}

\DeclareMathOperator{\vfin}{\mathcal{V}}

\def\als{\alpha_s}

\def\be{\begin{equation}}
\def\ee{\end{equation}}
\def\bea{\begin{eqnarray}}
\def\eea{\end{eqnarray}}

\newcommand{\pysecdec}{\texttt{pySecDec}{}}
\newcommand{\gosam}{\texttt{GoSam}{}}

\newcommand{\mzh}{m_{ZH}}
\newcommand{\mt}{m_t}
\newcommand{\mz}{m_Z}
\newcommand{\mh}{m_H}
\newcommand{\mw}{m_W}
\newcommand{\pth}{p_{T,H}}
\newcommand{\ptz}{p_{T,Z}}

\title{$ZH$ production in gluon fusion at NLO in QCD}
\author[a,i]{Long Chen,}
\author[b]{Joshua Davies,}
\author[c]{Gudrun Heinrich,}
\author[d]{Stephen P.~Jones,}
\author[c,e]{Matthias Kerner,}
\author[f]{Go Mishima,}
\author[g]{Johannes Schlenk,}
\author[h]{Matthias Steinhauser}

\affiliation[a]{Institute for Theoretical Particle Physics and Cosmology, RWTH Aachen University, 52056 Aachen, Germany}
\affiliation[b]{Department of Physics and Astronomy, University of Sussex, Brighton BN1 9QH, UK}
\affiliation[c]{Institute for Theoretical Physics, Karlsruhe Institute of Technology, 76128 Karlsruhe, Germany}
\affiliation[d]{Institute for Particle Physics Phenomenology, Durham University, Durham DH1 3LE, UK}
\affiliation[e]{Institute for Astroparticle Physics, Karlsruhe Institute of Technology, 76344 Eggenstein-Leopoldshafen, Germany}
\affiliation[f]{Department of Physics, Tohoku University, Sendai, 980-8578 Japan}
\affiliation[g]{Theory Group LTP, Paul Scherrer Institut, CH-5232 Villigen PSI, Switzerland}
\affiliation[h]{Institute for Theoretical Particle Physics, Karlsruhe Institute of Technology, 76128 Karlsruhe, Germany}
\affiliation[i]{School of Physics, Shandong University, Jinan, Shandong 250100, China}

\emailAdd{longchen@physik.rwth-aachen.de}
\emailAdd{j.o.davies@sussex.ac.uk}
\emailAdd{gudrun.heinrich@kit.edu}
\emailAdd{s.jones@cern.ch}
\emailAdd{matthias.kerner@kit.edu}
\emailAdd{go.mishima@icloud.com}
\emailAdd{johannes.schlenk@psi.ch}
\emailAdd{matthias.steinhauser@kit.edu}

\preprint{
  {\small 
    \hphantom{.}\hfill IPPP/22/19\\
    \hphantom{.}\hfill P3H-22-038\\
    \hphantom{.}\hfill KA-TP-08-2022\\
   \hphantom{.}\hfill TTP22-024\\
   \hphantom{.}\hfill TU-1147\\
   \hphantom{.}\hfill PSI-PR-22-08}\\
}

\abstract{
We present fully differential next-to-leading order results for Higgs production in association with a $Z$ boson in gluon fusion.
Our two-loop virtual contributions are evaluated numerically using sector decomposition, including full top-quark mass effects, and supplemented at high $p_T$ by an analytic high-energy expansion to order ($\mz^4, \mh^4, \mt^{32}$).
Using the expanded results we also present a study of the top-quark mass scheme uncertainty at large $p_T$.
}


\begin{document}

\maketitle

\section{Introduction}

\input{intro.tex}

\section{Setup of the calculation}
\label{sec:calculation}

\input{calculation.tex}

\section{Results}
\label{sec:results}

\input{results.tex}

\section{Conclusions}
\label{sec:conclusions}
\input{conclusions.tex}

\section*{Acknowledgements}
This research was supported by the Deutsche Forschungsgemeinschaft
(DFG, German Research Foundation) under grant  396021762 - TRR 257.
The work of GM was supported by JSPS KAKENHI (No. JP20J00328).
SJ is supported by a Royal Society University Research Fellowship (Grant URF/R1/201268).
The work of JD was supported by the Science and Technology Research Council (STFC) under the
Consolidated Grant ST/T00102X/1.
We also acknowledge support by the state of Baden-W\"urttemberg through bwHPC.

\bibliographystyle{JHEP}
\bibliography{combi_HE.bib}

\end{document}

%% file: intro.tex
Higgs boson production in association with a $Z$ boson is a
particularly interesting process as it probes both the Higgs boson
coupling to $Z$ bosons as well as to fermions.
Despite its relatively small cross section, the process $pp\to ZH$ in combination with the decay channel $H\to b\bar{b}$ was the ``discovery channel'' for
Higgs boson couplings to bottom quarks~\cite{ATLAS:2018kot,CMS:2018nsn}, as
while inclusive $H\to b\bar{b}$ suffers from large backgrounds,
the $Z$ boson offers straightforward triggering.
Recently, associated Higgs production was also used to place limits on the Higgs-charm coupling~\cite{ATLAS-CONF-2021-021,CMS:2022jed}.

The loop-induced gluon channel formally enters at next-to-next-to-leading order (NNLO), with respect to the $pp\to ZH$ process.
However, due to the dominance of the gluon parton distribution function (PDF) at the LHC, this channel is sizeable;
it contributes about 6\% to the total NNLO cross section and becomes significant in the boosted Higgs regime for $p_T^H\gtrsim 150$\,GeV~\cite{Altenkamp:2012sx,Harlander:2013mla,Englert:2013vua,Bizon:2021rww,Gauld:2021ule}.
For more details about calculations of higher-order corrections to the $pp\to ZH$ process we refer to Ref.~\cite{Heinrich:2020ybq}; here we focus on the gluon channel.

Experimental measurements of $ZH$ production~\cite{ATLAS:2019yhn,ATLAS:2020jwz,ATLAS:2020fcp,ATLAS:2021wqh,ATLAS:2022ers,CMS:2018vqh,CMS-PAS-HIG-19-005} still suffer from large statistical uncertainties, however the statistics will improve considerably in LHC Run 3 and at the High Luminosity LHC.
On the theoretical side, the scale uncertainties for this process are large.
They are dominated by the $gg\to ZH$ channel which, being loop induced, enters at its leading order (LO) into the simulation programs~\cite{Luisoni:2013cuh,Denner:2014cla,Hespel:2015zea,Goncalves:2015mfa,Campbell:2016jau,Granata:2017iod,Astill:2018ivh,Gritsan:2020pib}  used by the experimental collaborations.
Therefore next-to-leading order (NLO) QCD corrections to the $gg\to ZH$ process, calculated at LO in Ref.~\cite{Kniehl:1990iva}, are important; particularly so in view of providing constraints on anomalous couplings, because of the sensitivity of this process to both the Higgs couplings to fermions as well as to vector bosons. Furthermore, it provides a way to put constraints on the sign of the top quark Yukawa coupling as well as on its CP structure~\cite{Englert:2013vua,Goncalves:2015mfa,Hespel:2015zea,Freitas:2019hbk}, because the cross section has a component where this coupling enters linearly. Recently, it was shown~\cite{Yan:2021veo} that this process also has the potential to probe anomalous $Zb\bar{b}$ couplings and thus to shed light on the long-standing discrepancy of the forward-backward asymmetry $A_{FB}^b$ measured at LEP, with the Standard Model prediction.

The NLO QCD two-loop amplitude for $gg\to ZH$, including full top-quark mass effects, has been calculated numerically in Ref.~\cite{Chen:2020gae}.
The two-loop amplitude has also been calculated based on high-energy
(also sometimes called ``small-mass'')
and large-$\mt$ expansions, supplemented with
Pad{\'e} approximants to improve the description beyond the high-energy radius of convergence~\cite{Davies:2020drs}.
Results for the virtual corrections based on a transverse-momentum expansion are also available~\cite{Alasfar:2021ppe}.
Recently, the virtual corrections to both $gg\to HH$ and $gg\to ZH$, based on a combination of transverse-momentum expansion and high-energy expansion, have been presented in Ref.~\cite{Bellafronte:2022jmo}.
The total cross section and invariant-mass distribution for $gg\to ZH$ at NLO QCD has been calculated in Ref.~\cite{Wang:2021rxu}, based on a small-($\mz$, $\mh$) expansion~\cite{Wang:2020nnr} of the two-loop amplitude, retaining the full top-quark mass dependence.
An expansion of the virtual and real NLO contributions to $gg\to ZH$ for large $m_t$ has been computed in~\cite{Hasselhuhn:2016rqt}.

In this work we present full NLO QCD results for the $gg\to ZH$ process where the two-loop amplitude is based on a combination of the numerical results of Ref.~\cite{Chen:2020gae} and an extended version of the results from the high-energy expansion of Ref.~\cite{Davies:2020drs}, thereby providing reliable and accurate results in all kinematic regions, in particular in the boosted Higgs regime which is particularly sensitive to new physics effects.
A similar combination has already been carried out successfully for Higgs boson pair production~\cite{Davies:2019dfy}.
In addition, we consider two renormalisation schemes for the top quark mass, the on-shell (OS) scheme and the modified Minimal Subtraction (\msbar{}) scheme, and investigate how the scheme dependence impacts the phenomenological results.

This paper is organised as follows. In Section~\ref{sec:calculation} we describe the calculation of the NLO corrections and the combination procedure. Section~\ref{sec:results} contains results for the total cross section at different center-of-mass energies as well as $ZH$ invariant-mass and transverse-momentum distributions. The second part of Section~\ref{sec:results} is dedicated to the discussion of the top-quark mass scheme dependence, before we conclude in Section~\ref{sec:conclusions}.

%% file: calculation.tex
In this section we summarise the computation of the individual
contributions to the cross section at NLO and describe
the combination of the virtual corrections computed in~\cite{Davies:2020drs}
and \cite{Chen:2020gae}.

\subsection{\label{sub::virt}Virtual two-loop contributions}

\begin{figure}
  \centering
  \begin{subfigure}[b]{0.4\textwidth}
    \centering
    \includegraphics[width=\textwidth]{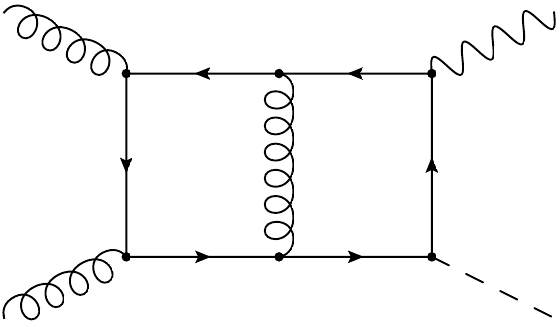}
  \end{subfigure}
  \qquad
  \begin{subfigure}[b]{0.4\textwidth}
    \centering
    \includegraphics[width=\textwidth]{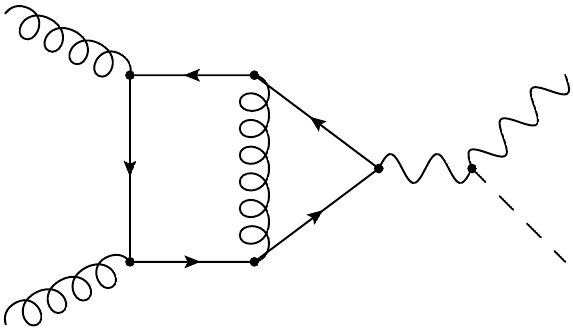}
  \end{subfigure}
  \caption{
  Representative Feynman diagrams for the virtual correction to the $ggZH$ amplitude.
  We neglect the masses of all quarks except the top quark; therefore, due to the Yukawa couplings, only the top quark gives a non-zero contribution for the box diagrams.
  All quark flavours contribute for the triangle diagrams, however the contribution from each massless generation is zero due to a cancellation between the up-type and down-type quarks.
  We calculate in the Feynman gauge and so also include the set of diagrams where the $Z$-boson propagators are replaced by Goldstone bosons.
  }
  \label{fig:virtuals}
\end{figure}

The calculation of the renormalised and infrared (IR) subtracted two-loop
amplitude, called $\vfin$,  is described in detail in
Refs.~\cite{Davies:2020drs} and~\cite{Chen:2020gae}. For completeness we
repeat in the following the most important steps.

In Ref.~\cite{Davies:2020drs} the amplitude of the process $gg\to ZH$ has been
written as a linear combination of six form factors. At one-loop order it is
straightforward to obtain exact results for the form factors.  At two-loops
expansions for large and small top-quark masses were performed. In this
work only the high-energy expansion, for which $\mh^2,\mz^2\ll \mt^2\ll s,|t|$, is of relevance. 
In Ref.~\cite{Davies:2020drs} an expansion up to order ($\mz^2, \mh^2, \mt^{32}$) was computed.
In this work we extend that result up to quartic order ($\mz^4, \mh^4, \mt^{32}$)  (including also the ``mixed'' quartic term $\mz^2 \mh^2$) and show that including
these quartic terms improves the agreement with numerical results.
\texttt{LiteRed}~\cite{Lee:2013mka} is used to expand the integrals appearing in the amplitude, followed
by an integration-by-parts (IBP) reduction to master integrals using \texttt{FIRE}~\cite{Smirnov:2019qkx}. The reduction relations
are substituted into the amplitude and simplified using \texttt{FORM}~\cite{Ruijl:2017dtg}.
In Ref.~\cite{Davies:2020drs} it was shown
that Pad\'e approximants for the expansion in $\mt$ significantly improve the description beyond the radius of
convergence of the naive expansion and reliable results can be obtained
for $p_T\ge 150$~GeV (see Fig.~\ref{fig:comb_pt}); we apply the same procedure here.
Note that in this approach one can construct the
high-energy expansion of $\mathcal{V}$ as a function of $\mh$, $\mz$ and $\mt$ which
allows for a variation of these parameters.  Furthermore, it is straightforward
to perform a scheme change and convert the top quark mass from the OS to the
\msbar scheme.  The subsequent construction of the new Pad\'e
approximants requires only negligible CPU time.

In Ref.~\cite{Chen:2020gae}, the two-loop amplitudes have been
calculated via a projection onto a basis of linear polarisation states as suggested in Ref.~\cite{Chen:2019wyb}.
For the IBP reduction of the resulting form factors to master integrals
we used \texttt{Kira}~\cite{Maierhofer:2017gsa,Klappert:2020nbg,Lange:2021edb}
in combination with the rational function interpolation library \texttt{FireFly}~\cite{Klappert:2019emp,Klappert:2020aqs}. 
For the numerical integration, using a quasi-finite basis~\cite{vonManteuffel:2014qoa} of master integrals is beneficial.
This basis is related to the default basis by dimension shifts and higher powers of propagators (dots).
For the derivation of the set of dimensional recurrence relations which connects integrals in $D+2n$ and $D$ dimensions, we used \texttt{LiteRed}~\cite{Lee:2013mka} and \texttt{Reduze}~\cite{vonManteuffel:2012np}.

To evaluate the master integrals, we applied sector decomposition as implemented in the program \pysecdec~\cite{Borowka:2017idc,Borowka:2018goh},  using a quasi--Monte Carlo algorithm~\cite{Li:2015foa,Borowka:2018goh} for the numerical integration. In particular, we made use of one of the new features of \pysecdec~\cite{Heinrich:2021dbf} to integrate a weighted sum of integrals such that the number of sampling points used for each integral is dynamically set, according to its contribution to the total uncertainty of the amplitude.

We renormalise the strong coupling in the \msbar scheme with 5 active quark flavours.
The top quark mass is renormalised in either the OS scheme or the \msbar scheme, as indicated.
Expanding each renormalised form factor $\mathcal{A}_{i=1,\ldots,n}$ in powers of the strong coupling, $a_s = \alpha_s/(4\pi)$, we may write
\begin{align}
\mathcal{A}_i^\mathrm{UV} = a_s A^{(0),\mathrm{UV}}_i + a_s^2 A^{(1),\mathrm{UV}}_i + \mathcal{O}(a_s^3)
\end{align}
and then obtain IR-finite amplitudes using the Catani-Seymour subtraction operator~\cite{Catani:1996vz}
\begin{align}
A^{(0),\mathrm{fin}}_i &= A^{(0),\mathrm{UV}}_i, \\
A^{(1),\mathrm{fin}}_i &= A^{(1),\mathrm{UV}}_i - I_1 A^{(0),\mathrm{UV}}_i.
\end{align}

The squared $2 \rightarrow 2$ amplitude, in the helicity basis, can be written as
\begin{align}
\overline{\sum} |A^\mathrm{fin}_i|^2 = a_s^2 \mathcal{B} + a_s^3 \mathcal{V} + \mathcal{O}(a_s^4),
\end{align}
where the squared Born amplitude ($\mathcal{B}$) and the Born-virtual interference ($\mathcal{V}$), are given by
\begin{align}
\mathcal{B} &= \overline{\sum} A^{(0),\mathrm{fin}}_i A^{*(0),\mathrm{fin}}_i, \\
\mathcal{V} &= \overline{\sum} \left( A^{(0),\mathrm{fin}}_i A^{*(1),\mathrm{fin}}_i + A^{(1),\mathrm{fin}}_i A^{*(0),\mathrm{fin}}_i \right).
\end{align}
The sum/average, denoted by $\overline{\sum}$, runs over all helicities and averages over the incoming spin and colour indices.

\subsubsection{Combination of the two approaches}

\begin{figure}
    \centering
    \includegraphics[width=0.6\textwidth]{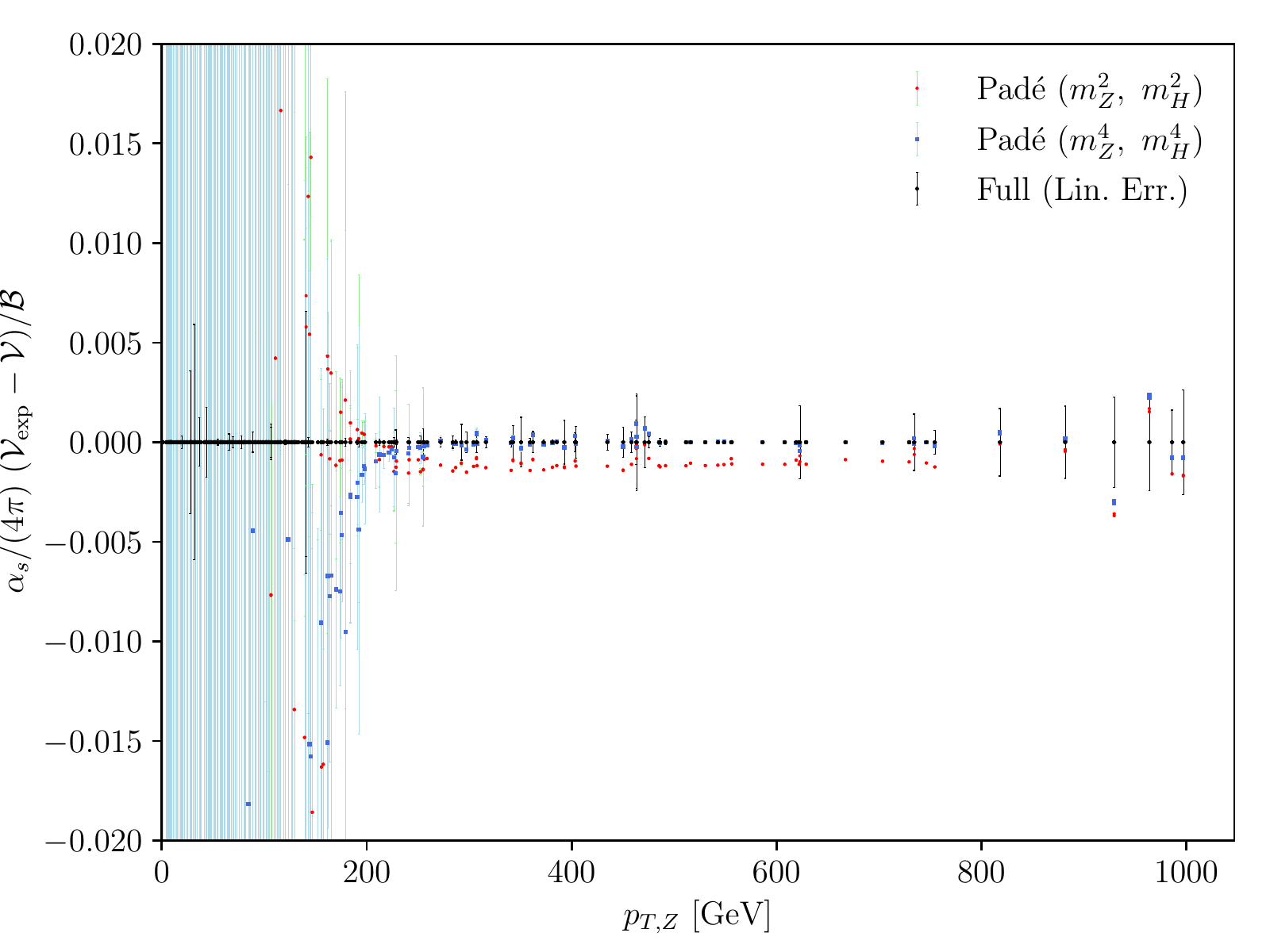}
  \caption{
  Comparison of the full and Pad\'{e} improved high-energy expanded results normalised to the Born result, as a function of $p_{T,Z}$.
  The Pad\'{e}-improved results are shown including up to quadratic terms, $(\mz^2,\ \mh^2)$, or quartic terms, $(\mz^4,\ \mh^4)$, in $\mz$ and $\mh$.
  For this comparison we fix the strong coupling constant to $\alpha_s=0.118$.
  }
  \label{fig:comb_pt}
\end{figure}

In Fig.~\ref{fig:comb_pt} we show the difference between the two calculations of the two-loop virtual amplitudes, relative to the LO amplitude. This difference is independent of the IR subtraction scheme used for removing the IR singularities. The plot shows that the Pad\'{e}-improved high-energy expansion converges on the full result for $p_T\gtrsim 150$~GeV.  However, using only the quadratic terms in $\mz^2$ and $\mh^2$ in the expansion, a difference at the two permill level remains even at large $p_T$. After including the quartic terms in the expansion, most of the points with $p_T>200$~GeV agree within the numerical uncertainty at the $2 \cdot 10^{-5}$ level (with a few outliers with large numerical uncertainty reaching up to the $2 \cdot 10^{-3}$ level). At low $p_T$ the differences increase, reaching up to 0.15\% at 200~GeV and up to 2.8\% at 150~GeV.

Since the two results are consistent for sufficiently large $p_T$, in the following we use the results based on the high-energy expansion for contributions with $p_T>150$~GeV and use the numerical evaluation of the amplitude only for $p_T<150$~GeV.  The two contributions are then combined at the histogram level. Our results are therefore valid in all phase-space regions, but avoid the costly numerical evaluation in large parts of the phase space.

\subsubsection{Phase-space sampling}

The integration of $\vfin$ over the phase space is achieved by a reweighting procedure based on the Born events. Specifically, we use the events of a LO calculation and apply the accept-reject method to obtain a list of sampling points for the virtual contribution  distributed according to the probability density function $\sim\,\mathcal L_{gg,0}\,|\mathcal B_0|^2\,\frac{d\mathrm{PS}}{f(p_T,m_{ZH})}$, where $\mathcal L_{gg,0}$ is the gluon-gluon luminosity as defined in Ref.~\cite{Borowka:2016ypz} and $\mathcal B_0$ is the LO matrix element as used in the LO calculation. The factor $d\mathrm{PS}$ is the Jacobian of the phase-space integration and the function $f(p_T,m_{ZH})$ can be used to enhance the number of sampling points in specific regions. Choosing, e.g., $f=f(m_{ZH})\propto \mathrm d\sigma_B/\mathrm d m_{ZH}$ leads to sampling points which are uniformly distributed in $m_{ZH}$, thus enhancing the number of events in the tail of the distribution, whereas $f(p_T,m_{ZH})=1$ results in sampling points distributed according to the fully differential LO cross section.

The virtual contribution to the cross section is then given by
\begin{align}
  \sigma_V = \frac{1}{N}\sum_{\text{N sampling points}} \frac{\mathcal L_{gg} \vfin}{\mathcal L_{gg,0}\,|\mathcal B_0|^2}\cdot{f(p_T,m_{ZH})}\,{\sigma_{B,0}},
\end{align}
where $\mathcal L_{gg}$ and $\vfin$ are the new gluon-gluon luminosity and virtual matrix elements, whereas $\mathcal L_{gg,0},\,\mathcal B_0$ and $\sigma_{B,0}$ are obtained from the original LO calculation, with the total LO cross section $\sigma_{B,0}$.

In the following results we use three different sets of sampling points, optimized for the total cross section, as well as the $m_{ZH}$ and $p_T$ distributions. While the exact form of $f(p_T, m_{ZH})$ is not important, a good choice can be obtained with a fit using a Pad\'{e} ansatz. In the region $p_T,m_{ZH}\ge 2$~TeV, where a uniform sampling of points is not needed, we keep $f$ constant.
In total, we use 1294 numerically evaluated  points distributed according to the differential LO cross section. For the $\mzh$ and $p_T$ distributions, we combine these results with sets containing an additional 6000 points, optimised for the corresponding distribution, evaluated using the Pad\'{e}-improved high-energy expansion.


\subsection{Computation of the real radiation contributions}

\begin{figure}
  \centering
  \begin{subfigure}[b]{0.3\textwidth}
    \centering
    \includegraphics[width=\textwidth]{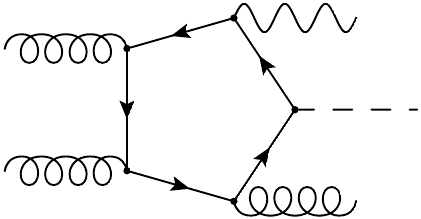}
  \end{subfigure}
  \hfill
  \begin{subfigure}[b]{0.3\textwidth}
    \centering
    \includegraphics[width=\textwidth]{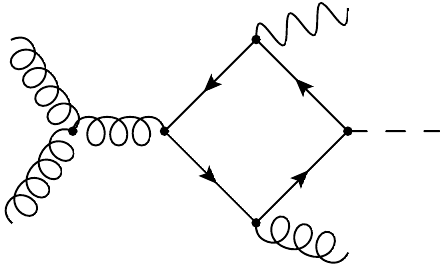}
  \end{subfigure}
    \hfill
    \begin{subfigure}[b]{0.3\textwidth}
    \centering
    \includegraphics[width=\textwidth]{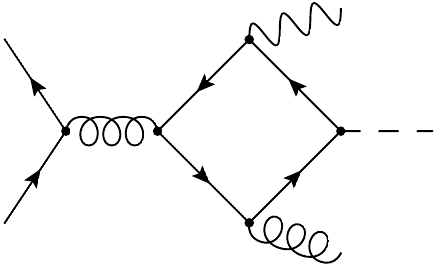}
  \end{subfigure}
  
    \begin{subfigure}[b]{0.3\textwidth}
    \centering
    \includegraphics[width=0.8\textwidth]{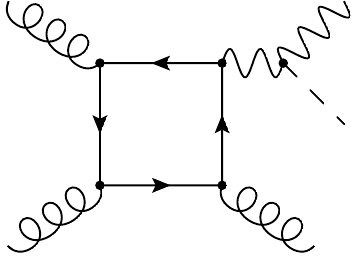}
  \end{subfigure}
  \hfill
    \begin{subfigure}[b]{0.3\textwidth}
    \centering
    \includegraphics[width=0.8\textwidth]{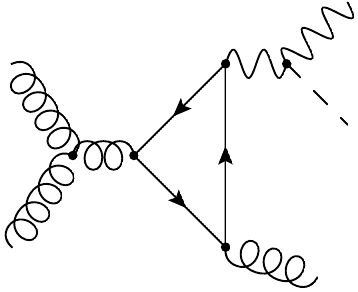}
  \end{subfigure}
  \hfill
    \begin{subfigure}[b]{0.3\textwidth}
    \centering
    \includegraphics[width=0.8\textwidth]{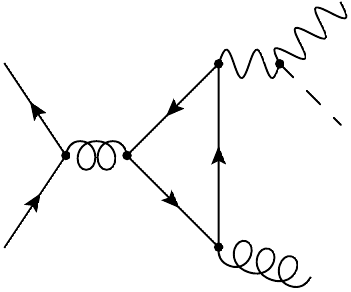}
  \end{subfigure}
  \caption{
  Representative Feynman diagrams for the real correction amplitudes $ggZHg$ and $q\bar{q}ZHg$, with $n_f = 5$ massless quarks and a massive top quark running in the closed fermion loops.
  We calculate in the Feynman gauge and so also include the set of diagrams in which the $Z$-boson propagators are replaced by Goldstone bosons.
  }
  \label{fig:reals}
\end{figure}

\begin{figure}
  \centering
    \begin{subfigure}[b]{0.3\textwidth}
    \centering
    \includegraphics[width=0.8\textwidth]{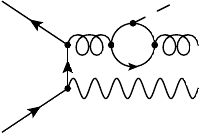}
  \end{subfigure}
  \qquad
    \begin{subfigure}[b]{0.3\textwidth}
    \centering
    \includegraphics[width=0.65\textwidth]{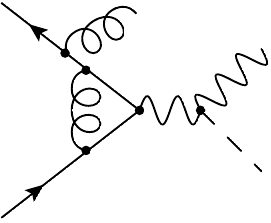}
  \end{subfigure}
  \caption{
  Representative Feynman diagrams for the class of real corrections excluded in this work;
  we exclude diagrams in which the $Z$ boson couples to the external quark line.
  }
  \label{fig:noreals}
\end{figure}

The real radiation matrix elements are calculated using the one-loop
amplitude generator \gosam~\cite{Cullen:2011ac,Cullen:2014yla}
together with an in-house \texttt{C++} code, similar to the one used in
Refs.~\cite{Borowka:2016ehy,Borowka:2016ypz},
where the IR singularities are subtracted in the Catani-Seymour
scheme~\cite{Catani:1996vz}, supplemented by a dipole phase-space
cut parameter $\alpha_\mathrm{cut}$~\cite{Nagy:2003tz}.
We have checked that our implementation of the dipoles reproduces the matrix element in the soft
and collinear limits and that our results are independent of 
$\alpha_\mathrm{cut}$ for $0.2 \le \alpha_\mathrm{cut} \le 1$.

To check the numerical precision of our real matrix elements we use several rotation tests (i.e., we perform azimuthal rotations about the beam axis and recompute the phase-space point).
We first compute the matrix element at a given phase-space point and a rotated phase-space point in double precision. If the results do not agree to 10 digits, we compute the phase-space point in quadruple precision and check if it agrees with the double-precision evaluations to 7 digits. If the results do not agree we compute a rotated point in quadruple precision and check that the quadruple-precision results agree to 10 digits; a vanishingly small fraction of points failed this test and were discarded.
Using the above procedure, we did not find it necessary to apply a technical cut to our $2 \rightarrow 3$ phase space.

In Fig.~\ref{fig:reals} we show examples of the Feynman diagrams included in our real radiation.
We include all diagrams appearing in the $ggZHg$ and $q\bar{q}ZHg$  amplitudes (as well as their crossings) which contain a closed fermion loop and have either a $Z$-boson or Goldstone boson coupled to that loop.
We consider $n_f = 5$ massless quarks and a massive top quark running in the fermion loops.
Due to the presence of the Yukawa coupling in the upper diagrams, the massless quarks give a non-zero contribution only for the lower row of diagrams. 
For the diagrams in the second and third column of the lower row, the contribution of each massless generation is zero due to a cancellation between the up-type and down-type quarks.
We calculate in the Feynman gauge and so also include the set of diagrams in which the $Z$-boson propagators are replaced by Goldstone bosons.

In Fig.~\ref{fig:noreals} we show examples of Feynman diagrams which are not included in this work; these diagrams do not have a $Z$ boson or Goldstone boson coupled to the closed fermion loop and so belong to the Drell-Yan class of diagrams, which we do not consider. The class of diagrams with the Higgs boson coupled to a closed quark loop, represented by the left figure, is UV/IR finite and separately gauge invariant; it was considered in detail in Ref.~\cite{Brein:2011vx}.

%% file: results.tex
\subsection{Total and differential cross sections}

Our results are based on a center-of-mass energy of $\sqrt{s}=14$\,TeV
unless stated otherwise. We use the NNPDF31\_nlo\_pdfas parton
distribution functions~\cite{NNPDF:2017mvq} with masses determined by the 
ratios $\mz^2/\mt^2 = 23/83$ and $\mh^2/\mt^2=12/23$ with $\mt=173.21$\,GeV and $\mw=80.379\,$GeV,
to 4 significant figures these ratios yield $\mz=91.18$\,GeV and $\mh=125.1$\,GeV.

Results for the total cross section with full top-quark mass dependence at three different center-of-mass energies, including scale uncertainties resulting from the 7-point scale variation, $\mu_{R,F} = \xi_{R,F}\, m_{ZH}$ with $\xi_{R,F}=(2,2),(2,1),(1,2),(1,1),(1,\frac{1}{2}),(\frac{1}{2},1),(\frac{1}{2},\frac{1}{2})$, are shown in Table~\ref{tab:totxs}.
Our result for $\sqrt{s}=13$\,TeV can be compared with that of Ref.~\cite{Wang:2021rxu}, where the virtual amplitude is computed in an expansion around small-$\mz$ and $\mh$, retaining the full $\mt$ dependence. 
At Born level, we observe that their result is $2.7\%$ larger than ours; we have verified that this is due only to the different choice of PDFs and masses ($\mz$, $\mh$ and $\mt$). 
At NLO their result is $2\%$ larger than ours, we ascribe this difference again to the different choice of PDFs and masses.
In Ref.~\cite{Wang:2021rxu} the scale uncertainty is assessed via a 3-point scale variation by a factor of 3; adopting this procedure we agree with their scale uncertainty of $^{+27\%}_{-21\%}$ at NLO.

\input{xs_table}

\begin{figure}
  \centering
  \begin{subfigure}[b]{0.49\textwidth}
    \centering
    \includegraphics[width=\textwidth]{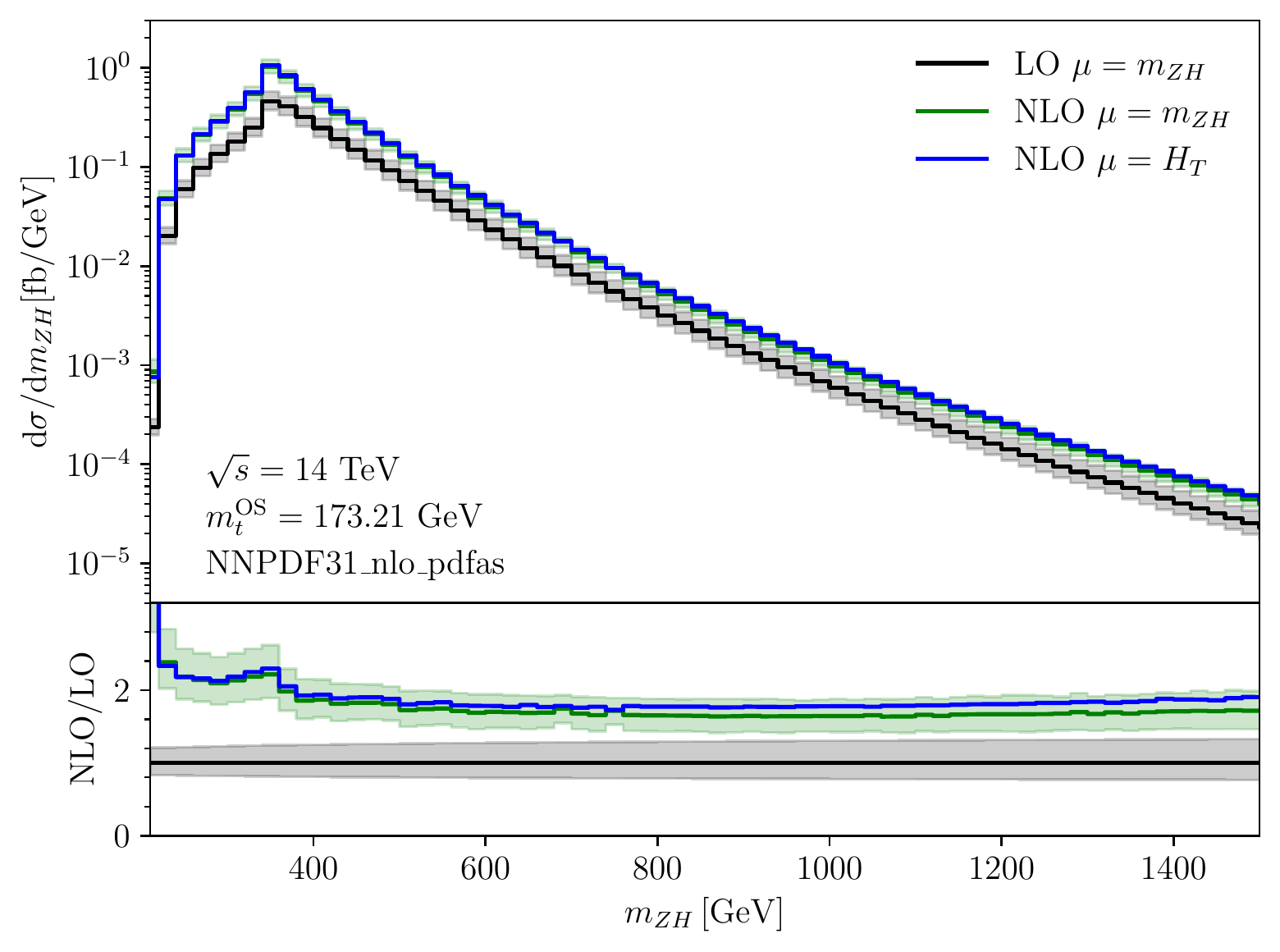}
  \end{subfigure}
  \hfill
  \begin{subfigure}[b]{0.49\textwidth}
    \centering
    \includegraphics[width=\textwidth]{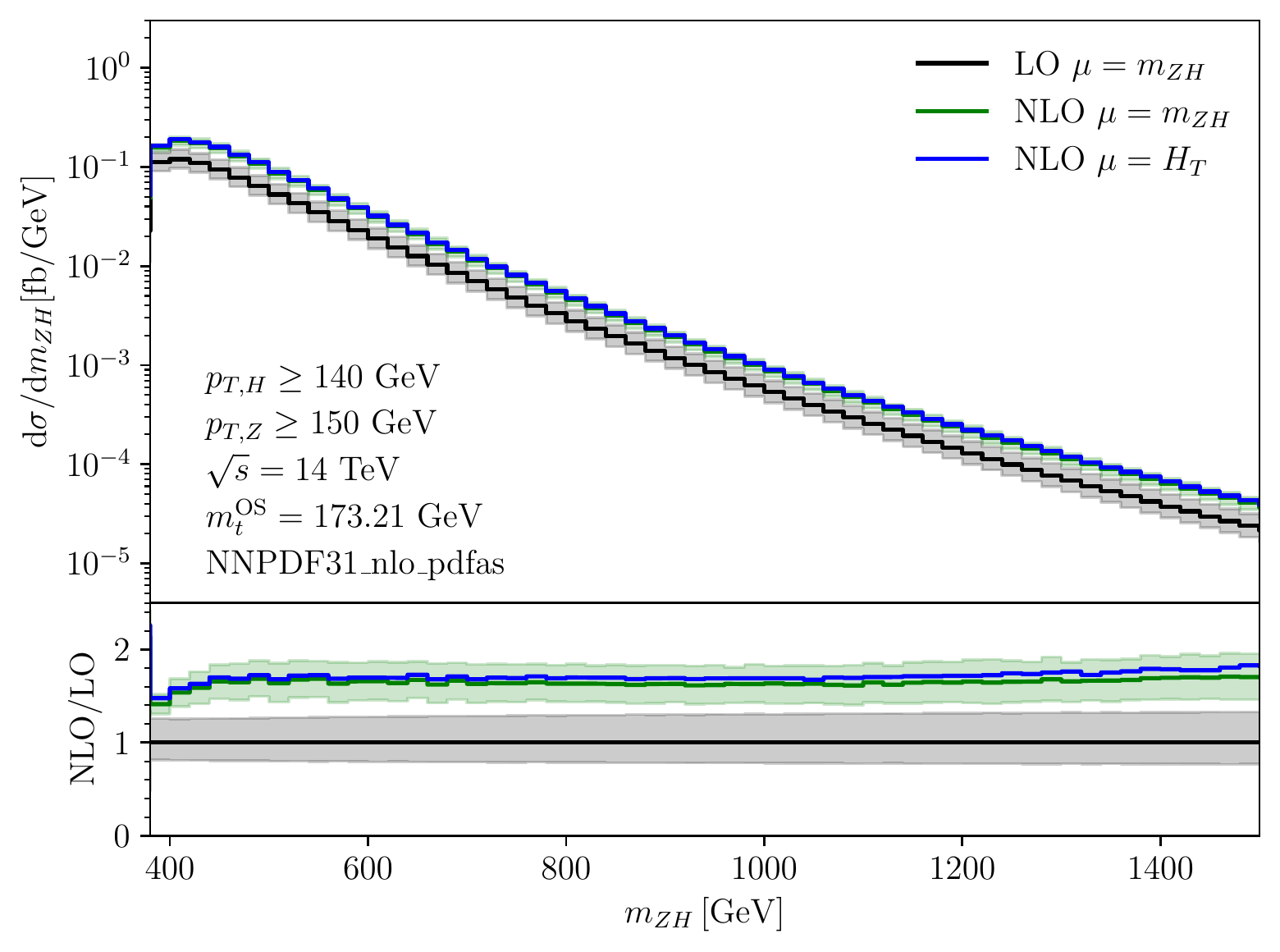}
  \end{subfigure}
  \caption{Invariant mass distribution at LO and NLO, including scale
    uncertainties resulting from a 7-point scale variation around the
    central scale $\mu=\mu_R = \mu_F = m_{ZH}$. 
    We also show NLO predictions for $\mu=H_T$.
    Left: fully inclusive,
    right: results based on $\pth \ge 140$\,GeV, $\ptz \ge 150$\,GeV.}
  \label{fig:mzh}
\end{figure}

Differential results for the invariant mass $\mzh=(p_Z+p_H)^2$ of the
$Z$-Higgs system are shown in Fig.~\ref{fig:mzh} for the central scale choices $\mzh$ and $H_T$, with
\begin{equation}
  H_T= \sum_{i=H,Z} \sqrt{m_i^2 + p_{T,i}^2} + \sum_k |p_{T,k}|,
  \end{equation}
where the sum runs over all final state massless partons $k$.
For the fully-inclusive case (left), the K-factor is relatively flat
with a value of about two, except at very low invariant masses where
threshold corrections are significant.
The kink in the distribution at $\mzh\simeq 350$\,GeV is related to the $t\bar{t}$-production threshold.
Only a small reduction of the scale uncertainty is observed going from LO to NLO.
Note that the quark-gluon channel for this process first opens up at
the NLO level.
The cuts $\pth \ge 140$\,GeV, $\ptz \ge 150$\,GeV (Fig.~\ref{fig:mzh}
(right)) somewhat decrease the K-factor.

\begin{figure}
  \centering
  \begin{subfigure}[b]{0.49\textwidth}
    \centering
    \includegraphics[width=\textwidth]{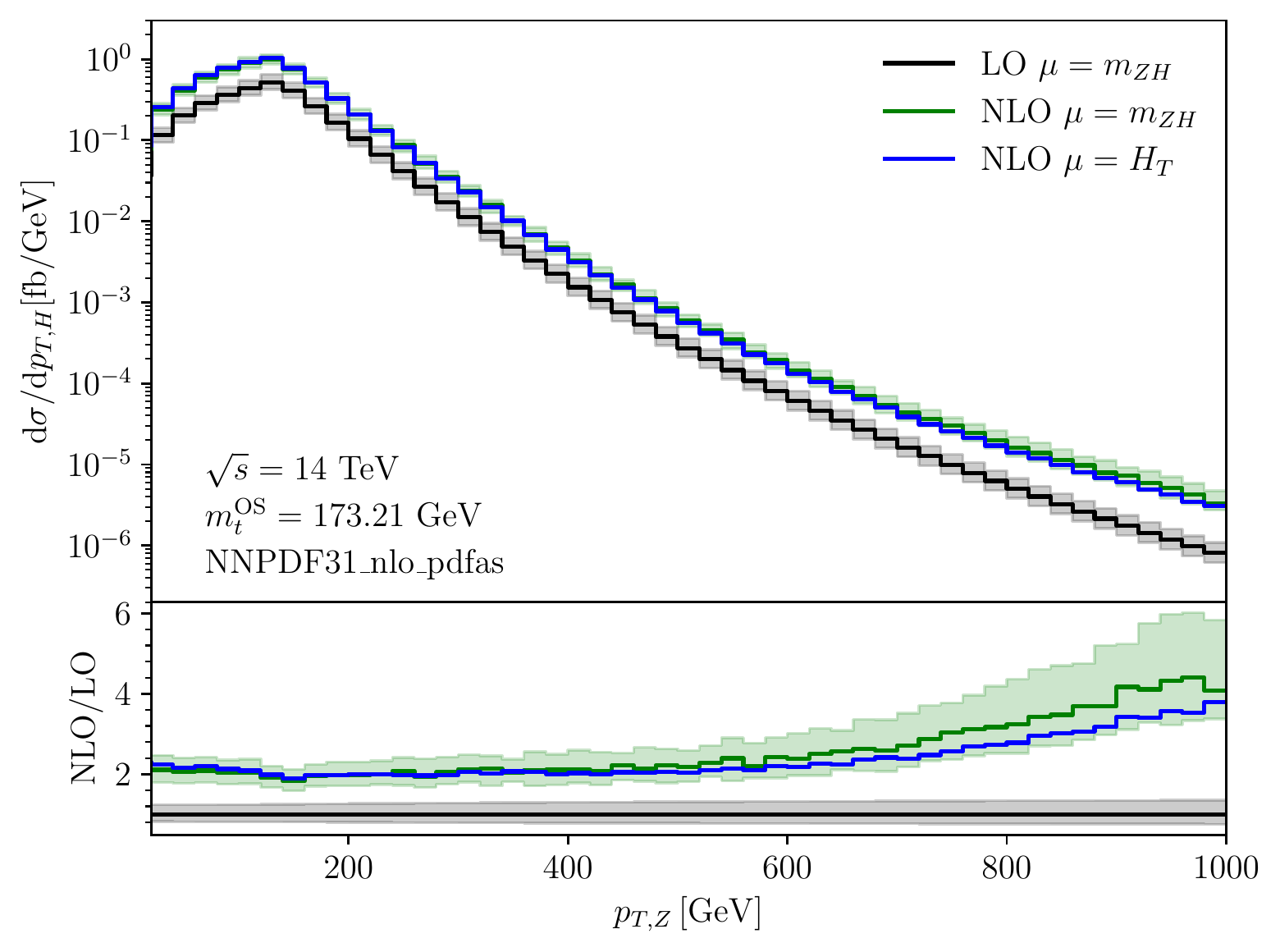}
  \end{subfigure}
  \hfill
  \begin{subfigure}[b]{0.49\textwidth}
    \centering
    \includegraphics[width=\textwidth]{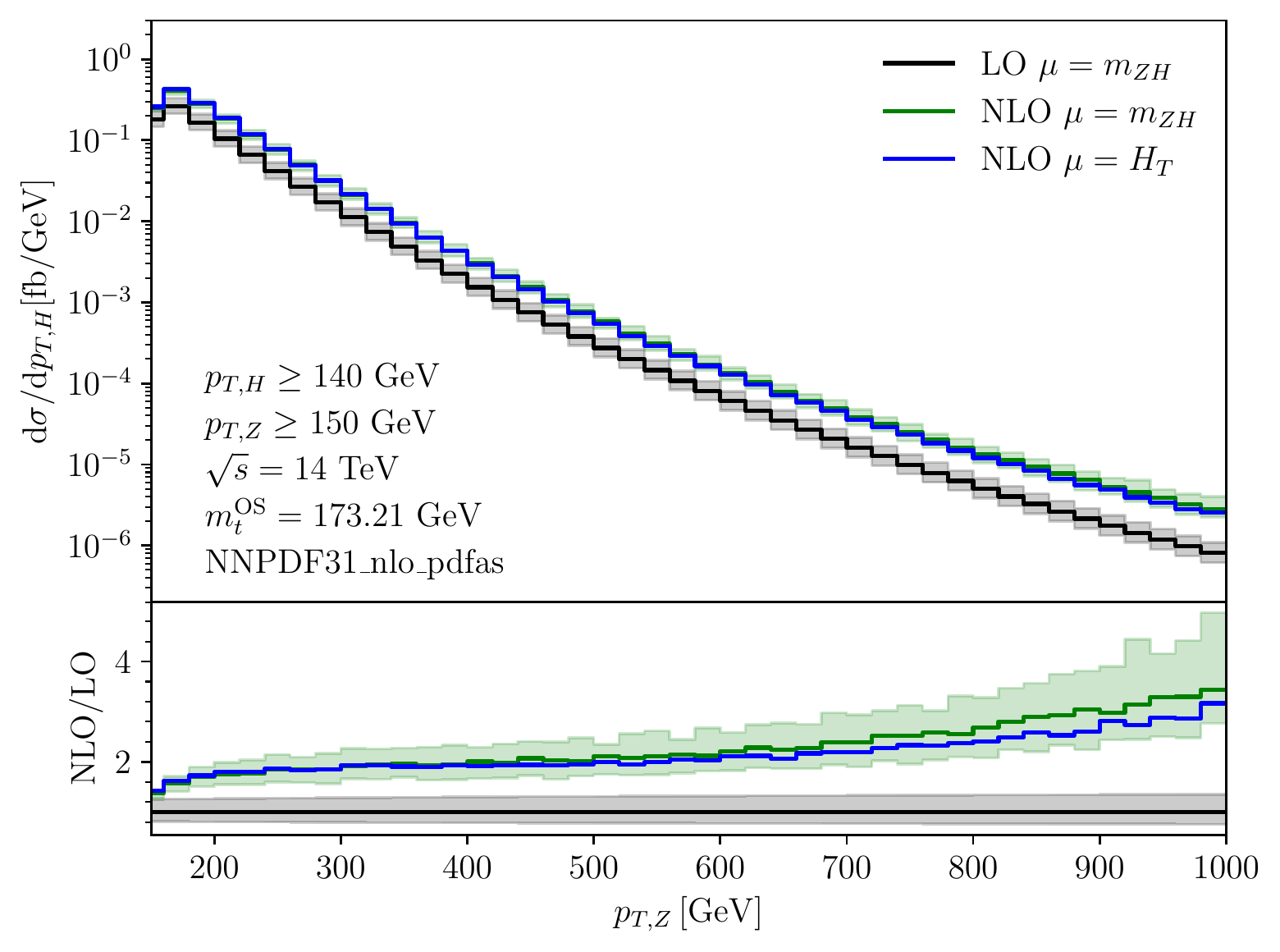}
  \end{subfigure}
  \caption{Distribution of the $Z$-boson transverse momenta at LO and NLO, including scale
    uncertainties resulting from a 7-point scale variation around the
    central scale $\mu=\mu_R = \mu_F = m_{ZH}$. 
    We also show NLO predictions for $\mu=H_T$.
    Left: fully inclusive,
    right: results based on $\pth \ge 140$\,GeV, $\ptz \ge 150$\,GeV.}
  \label{fig:mzh_ptz}
\end{figure}

\begin{figure}
  \centering
  \begin{subfigure}[b]{0.49\textwidth}
    \centering
    \includegraphics[width=\textwidth]{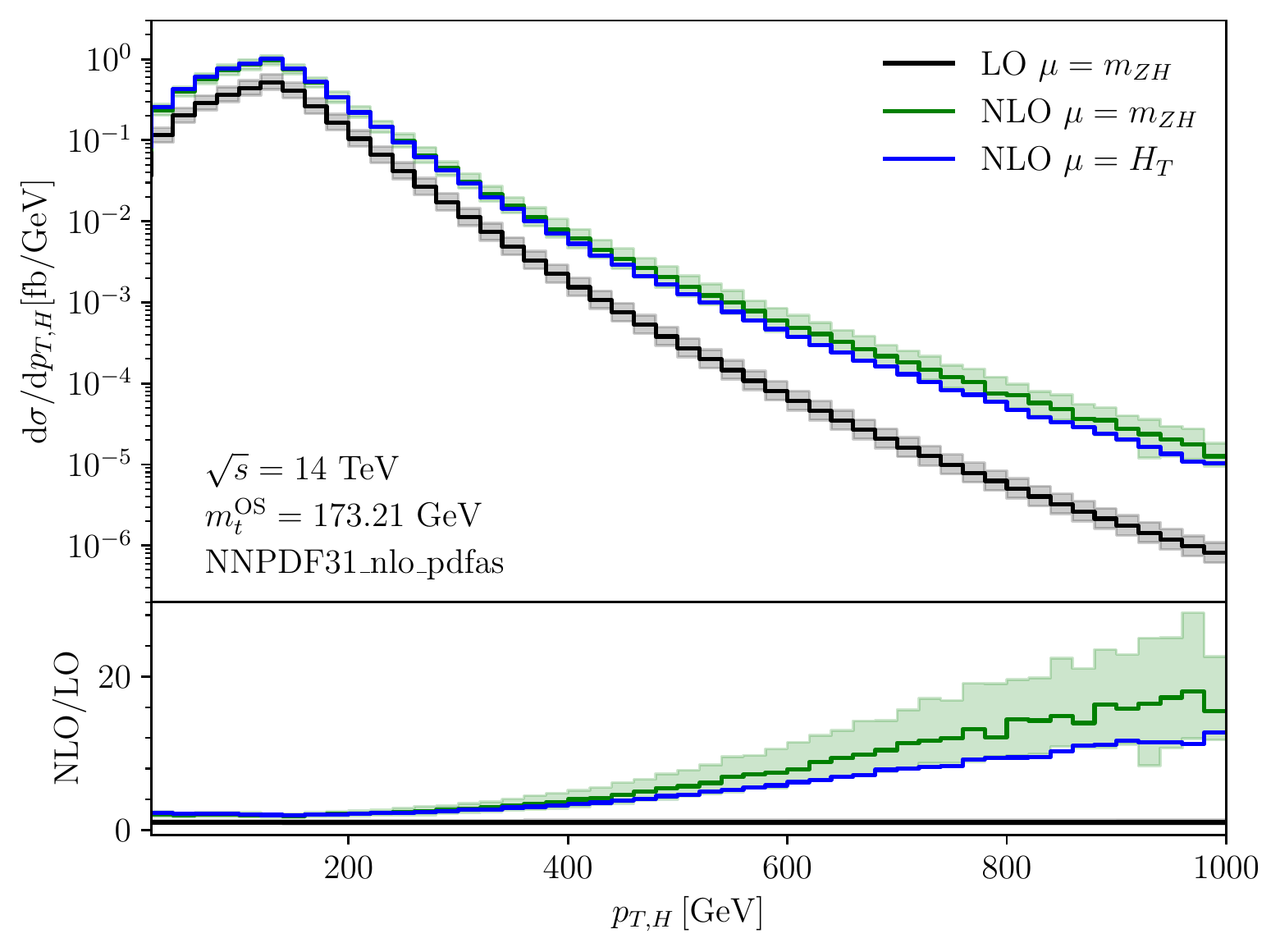}
  \end{subfigure}
  \hfill
  \begin{subfigure}[b]{0.49\textwidth}
    \centering
    \includegraphics[width=\textwidth]{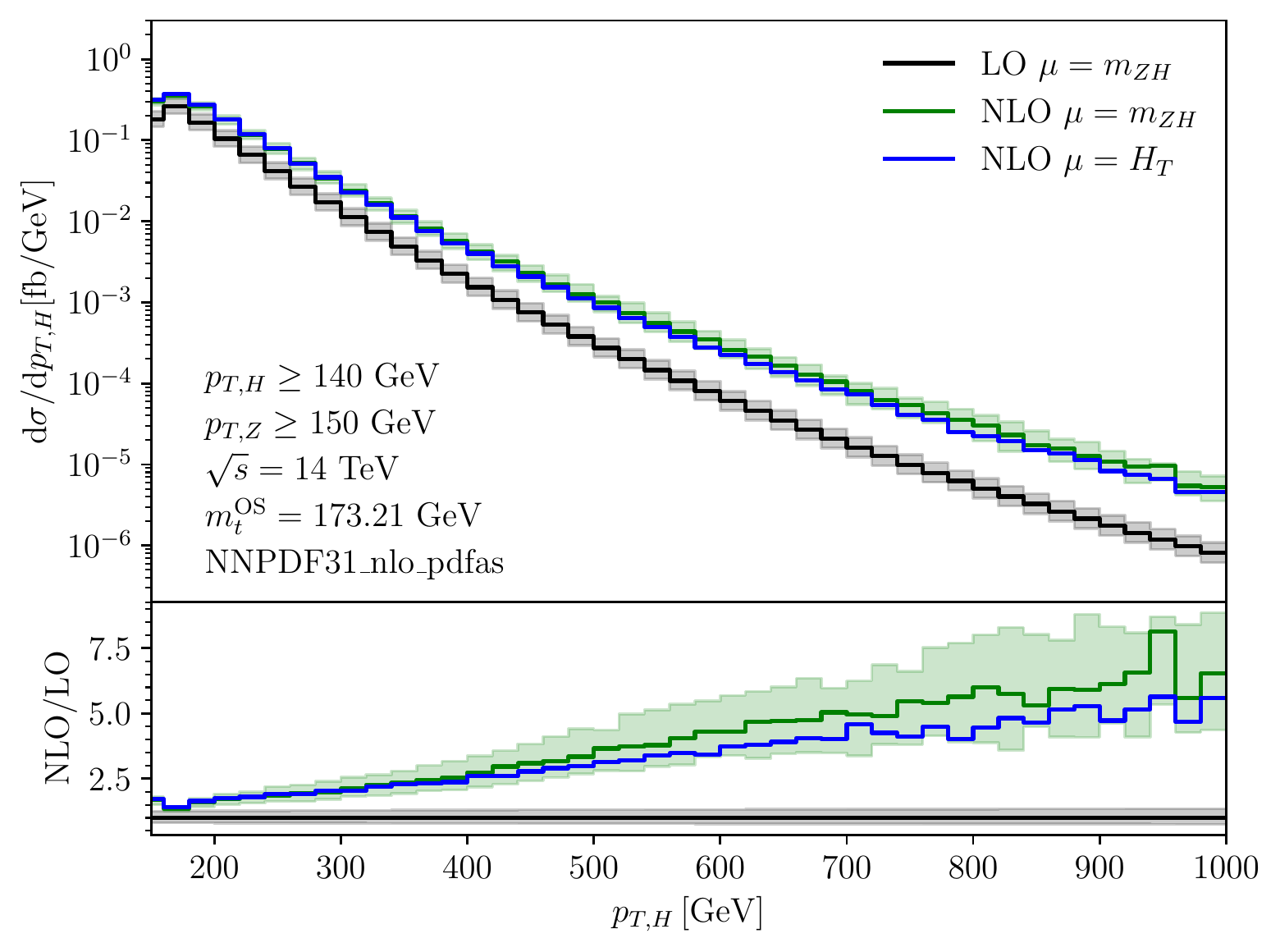}
  \end{subfigure}
  \caption{Higgs-boson transverse momentum distribution at LO and NLO, including scale
    uncertainties resulting from a 7-point scale variation around the
    central scale $\mu_R = \mu_F = m_{ZH}$. Left: fully inclusive,
    right: results based on $\pth \ge 140$\,GeV, $\ptz \ge 150$\,GeV.}
  \label{fig:mzh_pth}
\end{figure}

The $Z$-boson transverse momentum distributions at LO and NLO are
shown in Fig.~\ref{fig:mzh_ptz}.
In the left plot we observe a K-factor which rises with
increasing $\ptz$, reaching a value of almost 5 at $\ptz=1$\,TeV,
it  is only slightly tamed by the cuts on  $\pth$ and $\ptz$ (right plot).

Fig.~\ref{fig:mzh_pth} shows the Higgs-boson transverse momentum
distributions with and without $p_T$ cuts.
In the inclusive case (left) an extreme rise of the K-factor with
increasing $\pth$, up to values of about 20 towards $\pth=1$\,TeV, is
observed.
The cuts $\pth \ge 140$\,GeV, $\ptz \ge 150$\,GeV decrease this
K-factor by a factor of about 3 at large $\pth$ values.
The cuts have such a large effect on the K-factor of this distribution as they remove configurations
with a hard jet recoiling against a relatively hard Higgs while the $Z$ boson is soft,
this configuration dominates the tail of the distribution but is not present at LO.
This behaviour was already reported in Ref.~\cite{Hespel:2015zea}
and traced back to diagrams with $t$-channel gluon exchange,
it was further studied in Ref.~\cite{Amoroso:2020lgh}.
The reason why the rise of the K-factor is more pronounced in the
$\pth$ case than in the $\ptz$ case can be related to the coupling
structure of the $Z$ and Higgs bosons to top quarks.
In the diagrams where both the Higgs and the $Z$ boson are radiated from a top quark loop
, the probability to radiate a ``soft'' $Z$ boson while the Higgs boson recoils against a hard jet is related to the soft Eikonal factor $p^\mu/(p\cdot p_Z)$,
where $p^\mu$ generically denotes the 
radiator momentum.
The probability to radiate a ``soft'' Higgs boson on the other hand is proportional to $\mt/(p\cdot p_H)$.
The ratio of these Eikonal factors is $\simeq p_T/\mt\gg 1$,
thus at large transverse momentum $p_T$ of the radiator
it is more likely that the $Z$ boson is soft and the Higgs boson is hard.

\subsection{Investigation of different top quark mass renormalisation schemes}

We now turn to the discussion of the uncertainties stemming from the use of
different top quark mass renormalisation schemes.  Such uncertainties have
been investigated in detail for the case of Higgs boson pair production in
Refs.~\cite{Baglio:2018lrj,Baglio:2020ini,Baglio:2020wgt,Amoroso:2020lgh}. For
top quark pair production at NNLO, scheme uncertainties have been studied in
Ref.~\cite{Catani:2020tko}.
Top quark renormalisation scheme uncertainties also have been investigated for NLO $t\bar{t}H$~\cite{Martin:2021uqk} and $t\bar{t}j$~\cite{Alioli:2022lqo} production at the LHC, as well as for off-shell Higgs production and LO Higgs+jet production~\cite{Amoroso:2020lgh}.

\begin{figure}
  \centering
  \begin{subfigure}[b]{0.49\textwidth}
    \centering
    \includegraphics[width=\textwidth]{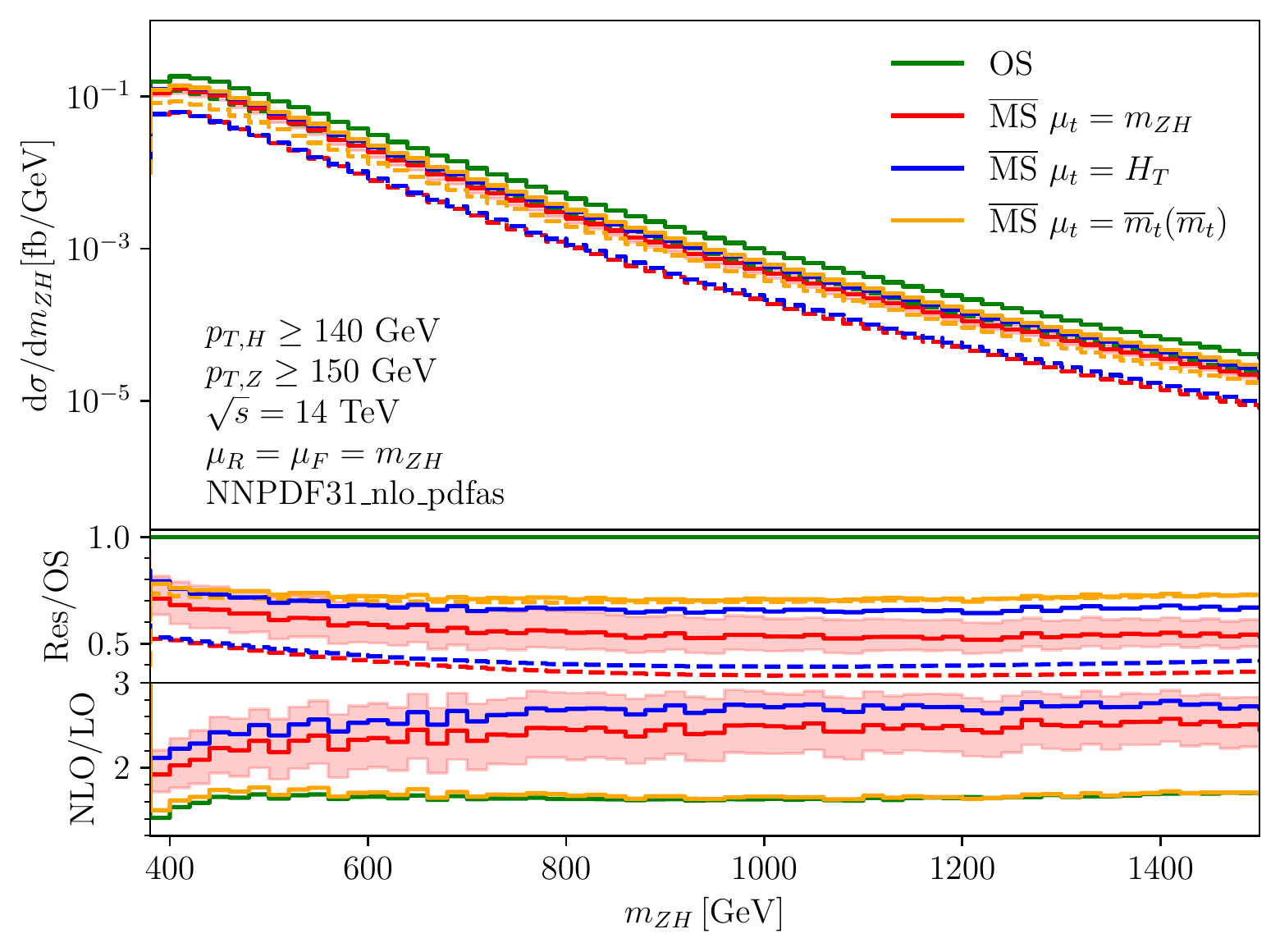}
  \end{subfigure}
  \hfill
  \begin{subfigure}[b]{0.49\textwidth}
    \centering
    \includegraphics[width=\textwidth]{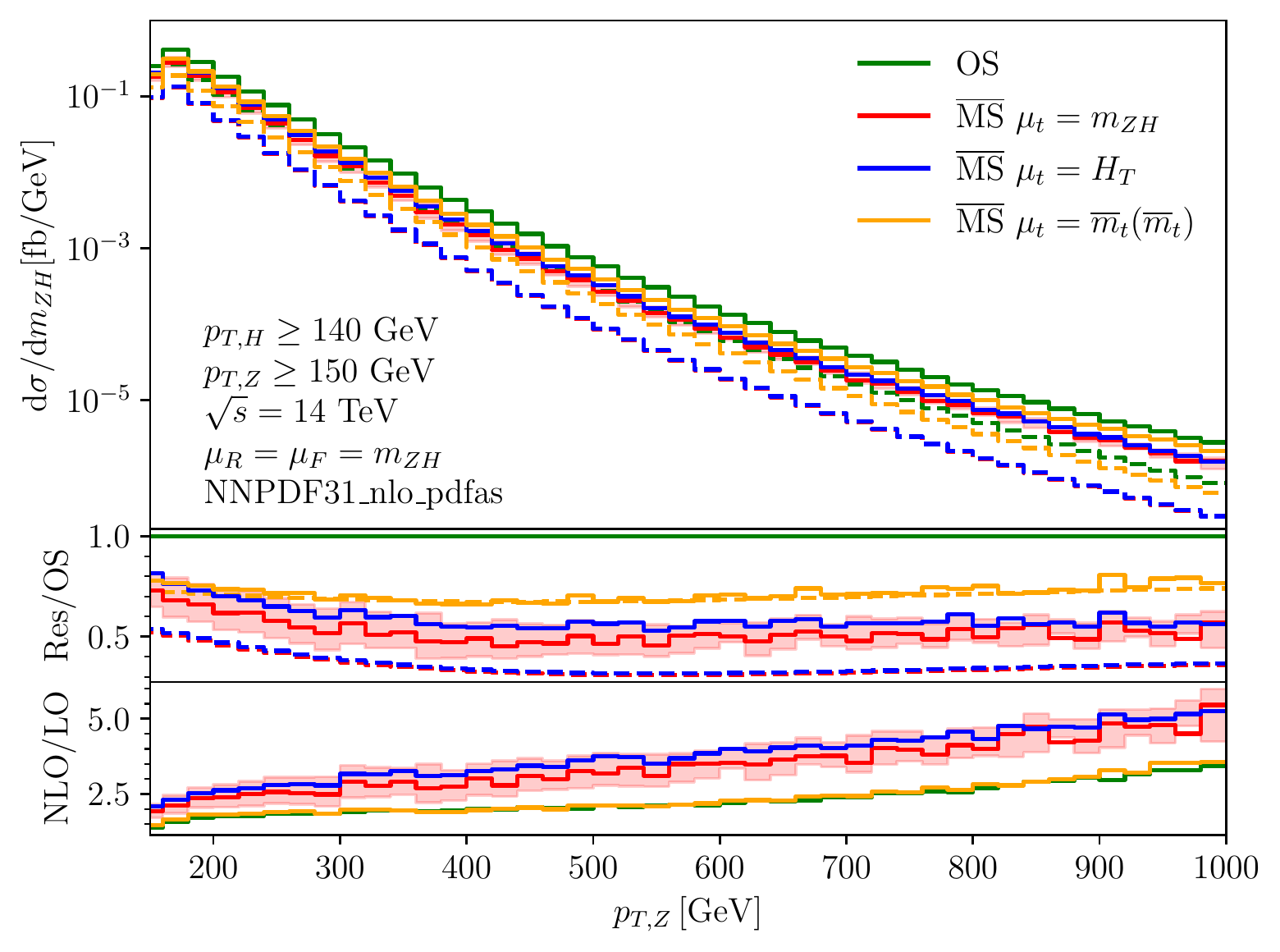}
  \end{subfigure}
  \caption{
  Comparison of differential results using various top quark mass schemes at LO (dashed) and NLO (solid).
  Left: Invariant mass.
  Right: $Z$-boson transverse momentum.
  }
  \label{fig:mt_scheme}
\end{figure}

In this section we investigate the top-quark mass renormalisation scheme dependence of $ZH$
production. For this purpose we convert the top quark mass to the
\msbar scheme, which is an appropriate renormalisation scheme in
the high-energy region. It is thus sufficient to perform the scheme change in
the analytic high-energy expansion of the virtual corrections, where
it is straightforward to obtain the corresponding analytic expressions
by making the replacement
\begin{equation}
	\label{eq:mtscheme}
	m_t \:\to\: \mtmsb(\mu_t)\: \left(1 + \frac{\als(\mu_R)}{4\pi} C_F \left\{4 + 3 \log\left[\frac{\mu_t^2}{\mtmsb(\mu_t)^2}\right] \right\} \right)
\end{equation}
and to apply the Pad\'e procedure as described in Section~\ref{sub::virt}.
Afterwards the result is combined with the real-radiation contributions
where the \msbar top quark mass is used in the numerical evaluation.

We make three different choices for the central value of the
top quark renormalisation scale, $\mu_t$,
\begin{itemize}
\item $\mu_t=m_{ZH}$
\item $\mu_t= H_T= \sum_{i=H,Z} \sqrt{m_i^2 + p_{T,i}^2} + \sum_k |p_{T,k}|\quad$ ($k$ sums over massless partons)
\item $\mu_t=\mtmsb(\mtmsb)$
\end{itemize}
and vary $\mu_t$ up and down by a factor of 2 to obtain an uncertainty
estimate.

For the conversion of the numerical values of the top quark mass between
the OS and the \msbar{} schemes we proceed as
follows: we first convert the top quark OS mass to the
\msbar scheme at the scale $\mu_t=\mtmsb$, at four-loop
accuracy. For our input values, $\mt=173.21$~GeV and $\alpha_s(\mz)=0.118$, this
gives $\mtmsb(\mtmsb) = 163.39$~GeV.  We then use the
renormalisation group equation, at five-loop accuracy with six active quark
flavours, to run from $\mu_t=\mtmsb$ to the desired renormalisation
scale for $\mtmsb$.
For both the numerical scheme conversion and the running we use the {\tt Mathematica}
and {\tt C++} codes~ {\tt RunDec} and {\tt
  CRunDec}~\cite{Chetyrkin:2000yt,Herren:2017osy}.
  
At LO the difference between the schemes is purely parametric and is driven both by the top quark mass appearing in the propagators and the Higgs-top Yukawa coupling.
At NLO the OS and \msbar schemes differ by the parametric choice of $m_t$ and a shift proportional to the derivative of the LO,
i.e., the mass counterterms, which partly compensates the parametric difference.

\input{mass_table}

In Fig.~\ref{fig:mt_scheme} we show predictions at LO (dashed lines) and NLO (solid lines) for $\mu_R = \mu_F = m_{ZH}$ with three different choices of the top-quark renormalisation scale, $\mu_t$.
The red band is generated by varying the scale $\mu_t = m_{ZH}$ up and down by a factor of 2, keeping $\mu_R = \mu_F = m_{ZH}$ fixed.
We observe that the scheme choice for the top quark mass has a large impact on the predictions for both the invariant mass distribution (with $p_{T,H} \ge 140\ \mathrm{GeV}$ and $p_{T,Z} \ge 150\ \mathrm{GeV}$ cuts) and the $p_{T,Z}$ distribution.
Focusing on the invariant mass distribution, we observe that at LO for $m_{ZH} \sim 1\ \mathrm{TeV}$ the OS result is approximately a factor of $2.9$ times the \msbar result with $\mu_t = m_{ZH}$.
At NLO the difference between the schemes is somewhat reduced, with the OS result around $1.9$ times the \msbar result with $\mu_t = m_{ZH}$, see Table~\ref{tab:mass}.
Taking, for example, the difference between the OS and the \msbar result with $\mu_t = m_{ZH}$ as a mass scheme uncertainty would result in a $^{+0\%}_{-65\%}$ uncertainty at LO and a $^{+0\%}_{-47\%}$ uncertainty at NLO for $m_{ZH} = 1\ \mathrm{TeV}$.
Alternatively, taking the \msbar result with $\mu_t = (m_{ZH}/2,\ m_{ZH},\ 2\, m_{ZH} )$ as an uncertainty gives $^{+26\%}_{-21\%}$ at LO and $^{+17\%}_{-14\%}$ at NLO for $m_{ZH} = 1\ \mathrm{TeV}$.
We observe a similar pattern for large $p_{T,Z}$, the difference between the OS scheme and the \msbar with $\mu_t = m_{ZH}$ scheme at $p_{T,Z} \sim 1\ \mathrm{TeV}$ is reduced from a factor of $2.8$ at LO to a factor of about $1.9$ at NLO.

The K-factor, defined as the ratio of the NLO result in a given scheme to the LO result in the same scheme, is typically larger in the \msbar scheme than in the OS scheme; this feature is also observed in Higgs pair production~\cite{Baglio:2020wgt}.
The K-factor of the invariant mass distribution is relatively flat for all scheme choices, with the dynamic scale choices $\mu_t = H_T$ and $m_{ZH}$ yielding $K \sim 2.5-2.7$ while the OS scheme has $K \sim 1.6$ for $m_{ZH} \sim 1\ \mathrm{TeV}$.
The \msbar scheme with $\mu_t=\mtmsb(\mtmsb)$,
where the logarithm appearing in Eq.~(\ref{eq:mtscheme}) vanishes,
has a very similar K-factor to the OS scheme; this differs from the $HH$ case where the two schemes had a similar shape but a different normalisation.
For the $p_{T,Z}$ distribution the pattern of K-factors for the different schemes is broadly the same as for the invariant mass distribution, but in all cases the K-factors rise with $p_{T,Z}$ reaching up to $\mathrm{K}=5$ for dynamic $\mu_t$ choices at $p_{T,Z} = 1\ \mathrm{TeV}$.

Comparing the results obtained here for $gg \rightarrow ZH$ to other loop-induced processes, such as off-shell Higgs production, Higgs pair production and Higgs plus jet production, we note that the $ZH$ process has a larger mass scheme dependence at LO.
For off-shell Higgs production and Higgs pair production going from LO to NLO approximately halves the uncertainty due to the mass scheme choice; in the $ZH$ case we also observe a reduction in the uncertainty, but by less than a factor of 2.

In the $HH$ case, in the high-energy limit, the triangle contribution is suppressed by a factor of 1/$s$ w.r.t.~the box form factors. Here, the leading high-energy behaviour of the box form factors has the form~\cite{Davies:2018qvx,Baglio:2020ini}
\begin{align}
    \label{eq:hhmtstructure}
	A_i^{(0)} \sim{}& m_t^2 f_i(s,t)\,\nonumber\\
	A_i^{(1)} \sim{}& 6 C_F A_i^{(0)} \log\left[\frac{m_t^2}{s}\right]\,,
\end{align}
where the $\log[m_t^2]$ term in $A_i^{(1)}$ is due to the renormalisation of $m_t$, and
the overall power of $m_t^2$ comes from the Yukawa couplings.
Converting to the \msbar scheme using Eq.~(\ref{eq:mtscheme}) results in a logarithm of the form $\log[\mu_t^2/s]$.
In Ref.~\cite{Baglio:2020ini} it was argued that choosing $\mu_t^2 \sim s$ minimizes these logarithms and is thus the
preferred central scale choice of the Yukawa couplings.
However, in the present
$ZH$ case, the structure is different.
Firstly, the triangle contribution is not suppressed w.r.t.~the box form factors, and secondly logarithms involving $m_t$ appear in the box form factors already at leading order.
Unlike in the $HH$ case, where the overall power of $m_t^2$ in Eq.~(\ref{eq:hhmtstructure}) comes entirely from the top Yukawa couplings, in $gg \rightarrow ZH$ one of the overall $m_t$ factors must come from the top-quark propagators, hence the leading term in the small-mass expansion is already power-suppressed by one power of $m_t$.
Similar, power-suppressed, mass logarithms have been studied in the context of single Higgs production, see for example Ref.~\cite{Liu:2021chn} and references therein.
The leading helicity amplitudes for $ZH$ in the high-energy limit have the form
\begin{align}
    \label{eq:zhmtstructure}
	A_i^{(0)} \sim{}& m_t^2 f_i(s,t) \log^2\left[\frac{m_t^2}{s}\right]\,,\nonumber\\
	A_i^{(1)} \sim{}& \frac{(C_A-C_F)}{6} A_i^{(0)} \log^2\left[\frac{m_t^2}{s}\right]\,,
\end{align}
Converting to \msbar generates terms of the form $C_F A^{(0)} \log[\mu_t^2/m_t^2]$,
therefore the choice $\mu_t^2=s$ does not eliminate the leading logarithms involving $m_t^2$, as is the case
for the box form factors of $HH$.

%% file: xs_table.tex
\begin{table}[h]
\centering
\begin{tabular}{ l  c  c }
\hline
\hline
$\sqrt{s}$ & LO [fb] & NLO [fb]\\
\hline
$13$ TeV		& $52.42^{+25.5\%}_{-19.3\%}$	& $103.8(3)^{+16.4\%}_{-13.9\%}$ \\
$13.6$ TeV	& $58.06^{+25.1\%}_{-19.0\%}$	& $114.7(3)^{+16.2\%}_{-13.7\%}$ \\
$14$ TeV		& $61.96^{+24.9\%}_{-18.9\%}$	& $122.2(3)^{+16.1\%}_{-13.6\%}$ \\
\hline
\hline
\end{tabular}
\caption{Total cross sections at LO and NLO with full top-quark mass dependence, evaluated at the scale $\mu_R = \mu_F = m_{ZH}$. 
The upper and lower values resulting from a 7-point scale variation are also shown.}
\label{tab:totxs}
\end{table}

%% file: mass_table.tex
\begin{table}[h]
\centering
\begin{tabular}{ l  c  c  c  c }
\hline
\hline
$m_t$ Scheme & LO [fb/GeV] & NLO [fb/GeV] & LO: OS/Res & NLO: OS/Res \\
\hline
$\mathrm{OS}$		& $5.35 \cdot 10^{-4}$	& $8.72(5) \cdot 10^{-4}$ & $1.0$ & $1.0$ \\
\msbar $(\mu_t=m_{ZH})$		& $1.87 \cdot 10^{-4}$	& $4.66(8) \cdot 10^{-4}$ & $2.85$ &  $1.87$ \\
\msbar $(\mu_t=H_{T})$	& $2.10 \cdot 10^{-4}$	& $5.69(8) \cdot 10^{-4}$ & $2.55$ &  $1.54$ \\
\msbar $(\mu_t=\mtmsb(\mtmsb))$	& $3.72 \cdot 10^{-4}$	& $6.18(6) \cdot 10^{-4}$ & $1.44$ &  $1.41$ \\
\hline
\hline
\end{tabular}
\caption{
Ratio of the on-shell result to results in various top quark mass schemes at LO and NLO for $\sqrt{s} = 14\ \mathrm{TeV}$ in the bin $m_{ZH} = [1000,1020]\ \mathrm{GeV}$.
Transverse momentum cuts of $p_{T,H} \ge 140\ \mathrm{GeV}$ and $p_{T,Z} \ge 150\ \mathrm{GeV}$ are applied and the remaining scales are set to $\mu_R = \mu_F = m_{ZH}$.}
\label{tab:mass}
\end{table}

%% file: conclusions.tex
The $gg \rightarrow ZH$ channel contributes to the $pp \rightarrow ZH$ process starting at NNLO, accounting for around $6\%$ of the total cross section.
However, the gluon-fusion channel suffers from a large scale dependence at LO and is a significant source of theoretical uncertainty for $Z$ boson production in association with a Higgs boson at the LHC~\cite{ATLAS:2018kot,CMS:2018nsn,CMS:2018vqh,ATLAS:2020jwz,ATLAS:2020fcp,ATLAS:2021wqh,ATLAS:2022ers}.

In this work, we have presented the complete NLO corrections for the loop-induced gluon-fusion channel; they increase the gluon-fusion cross section by about a factor of 2, and reduce the scale dependence.
We have investigated the invariant mass distribution and transverse momentum distributions for both the $Z$ boson and Higgs boson.
At large transverse momentum, we found that the NLO corrections can be very large, more than 10 times the LO result for $p_{T,H}$; the origin of this behaviour can be traced back to extremely large real radiation corrections when a soft $Z$ boson is radiated from a top quark loop~\cite{Hespel:2015zea}.

We have also studied the top quark mass scheme uncertainties for this channel, i.e., the difference between results produced with the top quark mass renormalised in the on-shell scheme and the $\overline{\mathrm{MS}}$ scheme.
As for other loop-induced processes with a scale above the top quark pair-production threshold~\cite{Baglio:2018lrj,Baglio:2020ini,Baglio:2020wgt,Amoroso:2020lgh}, we found a large mass scheme uncertainty at LO.
At NLO the mass scheme uncertainty is smaller than at LO, but remains at least as large as the usual renormalisation and factorisation scale uncertainties.

The inclusion of the NLO corrections to the gluon-fusion channel is essential for correctly describing $ZH$ production at the LHC and HL-LHC.
The size of the NLO corrections, especially for the transverse momentum distributions, and the mass scheme uncertainty motivate a calculation of $gg \rightarrow ZH$ beyond NLO and the study of this process beyond fixed order.